\documentclass[]{emulateapj}
\usepackage{amsmath}

\lefthead{Wada}
\righthead{Galactic Shock}

\newcommand{\bea}{\begin{eqnarray} }
\newcommand{\eea}{\end{eqnarray}}
\def\mbf#1{\mbox{\boldmath ${#1}$}}

\begin{document}	

%\title{Galactic Whirlpool in Chaos}
\title{Instabilities of Spiral Shocks -- II. 
A quasi-steady State in the
multi-phase inhomogeneous ISM}
\author{Keiichi Wada$^1$}
\affil{National Astronomical Observatory of Japan, Mitaka, Tokyo
181-8588, Japan\\
E-mail: wada.keiichi@nao.ac.jp}
\altaffiltext{1}{Department of Astronomical Science, The Graduate University for Advanced Studies, Osawa 2-21-1, Mitaka, Tokyo 181-8588, Japan.}

\begin{abstract}
The ``galactic shocks'' \citep{fujimoto68,roberts69} is investigated
using a full three-dimensional hydrodynamic simulations, taking into account
self-gravity of the ISM, radiative cooling, and star formation followed
by energy feedback from supernovae. This is an essential progress from
the previous numerical models, in which 2-D isothermal,
non-self-gravitating gas is assumed.  We find that the classic galactic
shocks appears is unstable and transient, and it shifts to a globally
quasi-steady, inhomogeneous pattern due to non-linear development of
instabilities in the disk. The spiral patterns consists of many GMC-like
dense condensations, but those local structures are not steady, and they
evolves into irregular spurs in the inter-arm regions.
%The quasi-steady spiral arms are sustained
%in a stochastic fashion in the multi-phase interstellar medium. 
 Energy
feedback from supernovae do not destroy the quasi-steady spiral arms,
but it mainly contributes to vertical motion and structures of the ISM.
The results and methods presented here are a starting point for 
more consistent treatment of the ISM in spiral galaxies, in which
effects of magnetic fields, radiative transfer, chemistry, 
and dynamical evolution of a stellar disk are taken into account.

%This results suggest that magnetic fields are not necessary to
%sustain inhomogeneous spiral structures on kpc scales.
%Detailed  molecular gas structures on a 10 parsec scale will be
%revealed by the next-generation radio interferometer ALMA in

\end{abstract}
% history


\section{INTRODUCTION}
A standing spiral shock solution was first discovered in the 60s
numerically in a rotating gas disk in the galactic potential with a
tightly wound spiral perturbation (i.e. the ``pitch angle'' is very
small). Since then in the most of theoretical studies on the galactic
``shock'' \citep[e.g.][]{fujimoto68,roberts69,wood75,johns86,lubow86},
 the interstellar medium (ISM) is treated as a
isothermal and homogeneous fluid with
two-dimensional approximation, or it is modeled as a $N-$body system of
small cloudlets \citep[e.g.][]{tomisaka86}.  
%It was proved that a spiral shock, in which the gas is
%compressed by about 100 times, is formed, therefore a star formation is
%expected at the downstream of the shock due to gravitational instability
%of the compressed layer.
% Self-gravity of the ISM and its
%discrete nature were studied later on [Tomisaka 1987]. The N-body
%simulations show that Giant molecular clouds (GMC) are formed through
%collisional growth of smaller clouds under the influence of a spiral
%potential. 
Global evolution of the spiral shock was studied in the 80s
\citep{johns86} using time-dependent, two-dimensional (2D) hydrodynamic
simulations show that spiral shocks are stable and long-standing for
various pitch angles.
% But, an isothermal, non-self-gravitating gas
%disk was still assumed.
%These classic studies do not consistently explain 
%the complicated structures of the real ISM in spiral galaxies,
%such as molecular clouds, 
%The galactic shock caused by stellar spiral density
%waves followed by star formation in GMCs then become a standard theory
%to explain the long-lived spiral structures.

However, the steady, smoothed galactic shock does not consistently
explain the complicated distribution of the ISM around spiral arms and
the inter-arm substructures akin to so-called `spurs' or
`feathers'\footnote{In this paper, we refer the terms 'spurs' or
'feathers' as inter-arm gas structures, which often show quasi-periodic
features associated with main spiral arms. See Paper I,
\citet{shetty06}, and \citet{kim02} for numerical examples, and \citet{vigne06} for
observations. On the other hand, `blanches' are sub-structures
bifurcated from main spiral arms.  They are smoother and longer
azimuthal extent than spurs, which may be caused by resonances
\citep{chark03}.} observed in real spiral galaxies
\citep{elm80,scov01,calz05,vigne06}. Moreover, observed molecular clouds
in galactic disks do not match the picture of hydrodynamic shocks in a
uniform media. 
%Using two-dimensional (2D) hydrodynamic, and
%magneto-hydrodynamic simulations, it was suggested that the spurs are
%caused due to gravitational \citep{shetty06} or magneto-Jeans
%instability \citep{kim02}.  Based on a similar analytical approach as
%\citet{roberts69}, \cite{shu73} pointed out that azimuthally
%non-sinusoidal features could appear due to ultra-harmonic
%resonances. This was confirmed by 2D hydrodynamic simulations by
%\citet{chark03}, in which spiral shocks bifurcate. This substructure
%looks similar to so-called ``blanch''.
%simulations of a local patch of gaseous disk with a spiral potential
%were performed, and it is claimed that prominent spurs are formed due
%to. 
It was however shown by full 2D global simulations of a
non-self-gravitating, isothermal gas disk that the spurs are in fact
natural consequences of ``wiggle'' instability, which is caused by a
purely hydrodynamic phenomenon, i.e. Kelvin-Helmholtz instability
\citep{wadakoda04} (hereafter Paper I).  This phenomenon was also found in
\citet{shetty06}. They also pointed out that the features only grow in
the inner-most several kpc regions if self-gravity and magnetic fields
are ignored.

%--- cut in 3rd revision --- 
%changing the pattern speed of the spiral potential does not
%dramatically alter the spiral structure, implying that
%resonances are not a direct cause of the spurs (see also 
%discussion in \S 4).

%A relevant numerical scheme with a high spatial resolution is essential
%to resolve the unstable nature of the shock.
More recently, three-dimensional (3D) response of the gas to the spiral potential was
modeled using a local shearing box approximation in isothermal, MHD
simulations taking into account self-gravity \citep{kim06},
and found that the wiggle instability is suppressed by radial flapping
motion of the shock.
% if the vertical structure of the gas disk is considered.

%This somewhat confusing situation concerning the stability and structure
%of the galactic shock now should be resolved by a more realistic,
%comprehensive model of the ISM in galactic disks, that is

These previous results suggest that hydrodynamic effects, self-gravity
of the gas and magnetic fields play some important roles on the gas
structures in a spiral potential. 
%Gravitational interactions between
%the stellar potential and the gas would be also important. 
However, an important feature of the real ISM has been ignored; its
inhomogeneous multi-phase structures.
Apparently the ISM is not `isothermal fluid', nevertheless
it is assumed in most HD and MHD simulations\footnote{\citet{dobbs07}
recently studied gas dynamics in a spiral potential, taking into account
multi-phase nature of the ISM.  However, in their non-self-gravitating
SPH simulations, an energy equation with realistic cooling and heating
processes is not solved, alternatively warm ($10^4$ K) and cold ($100$
K) components are treated separately as isothermal gases without phase
exchange.}. Effects of self-gravity and magnetic fields highly depend on
gaseous temperatures and phases. In this sense, an energy equation with
a realistic cooling and heating processes should be solved before taking
into account those effects.  Local box-approximation with a shearing
periodic boundary \citep[e.g.][]{kim06} is not necessarily relevant for
representing dynamics of the multi-phase ISM in galactic disks, because typical scales
of the inhomogeneous structures of the ISM are not small enough compared
to the disk size \citep[cf.][]{wada01, wada07, tasker06}.  Moreover, since
gravity is a long-range force, global, 3D simulations for the whole disk
are essential if self-gravity of the gas is considered.  This is also necessary for
the multi-phase ISM, because the scale height and velocity dispersion
are different for cold and hot gases.  In 3D hydrodynamic simulations of
self-gravitating gas disks with the radiative cooling, we should take
into account energy feedback from stars, otherwise the cold gas disk
becomes unrealistically thin and the gravitational instability is
strongly affected.

%The 2D models should be essentially used 
%The isothermal approximation is not suitable for the real ISM,
%especially for the self-gravitating model, since the gravitational
%instability strongly depends on temperature of the ISM.
%The ISM is multi-phase with wide ranges of density and temperature, and it is 
%essential to study how the GMCs are formed under the influence of
%the spiral potential.

Here we show, for the first time, 3D evolution of the ISM in a galactic
spiral potential, taking into account realistic radiative cooling and
energy feedback from stars, especially from supernovae explosions.  The
simulations are performed for the whole gas disks without any
assumptions for symmetry.  In order to ensure a high spatial resolution
(10 pc), which is essential to reproduce the multi-phase ISM, we here
focus on a central part of a relatively small galaxy (radius is 2.56 kpc
and the maximum circular velocity is 150 km s$^{-1}$). However, this is
large enough to see the global effect of the spiral potential on the
inhomogeneous ISM. Our simulations clarify apparent discrepancy between the
steady solution of ``galactic shock'' and the complicated, non-steady
structures of the ISM in real spiral galaxies.  This is a major progress
from the previous simulations, and it will be a starting point for more
realistic numerical models of the ISM, taking into account `lived'
stellar potential, magnetic fields, UV radiation from massive stars, and
chemistry of molecules and atoms.

%the true structure and stability of the ``galactic shock''.
%For the comparison, we also show higher resolution 2D simulations
%and 3-D result with isothermal EOS.
%The ISM is initially distributed in a disk with a radius of 2.5 kpc (3D
%model) or 10 kpc (2D model).
%In the 3-D model, massive ``stars'' are assumed to be 
%formed in a cold ($T< 100$ K) and dense ($n > 600$ cm$^{-3}$)
%media,  which are formed by gravitational and thermal instability of the
%gas under influence of a fixed spiral potential (pitch angle is 10 or 15 deg).
%The stars orbiting in 
%the galactic potential 
%are exploded as type-II supernovae in 1-10 Myr after they are
%born, and energy of a supernovae is injected into a grid-cell (10 pc).

\section{METHODS}
Evolution of a gas disk (total mass is $4.3\times 10^9 M_\odot$, radius
is 2.5 kpc) in a fixed gravitational potential $\Phi_{\rm ext}$ is
solved by an Eulerian hydrodynamic method with a uniform Cartesian
grid \citep{wada01, wada01b,wada07}.  Here we briefly summarize the
numerical scheme.  \setcounter{footnote}{0} We solve the following
conservation equations and Poisson equation in three dimensions using
$512\times 512 \times 64$ grid points with 10 pc resolution.
  \begin{eqnarray}
\frac{\partial \rho}{\partial t} &+& \nabla \cdot (\rho \mbf{v}) = 0,
\label{eqn: rho} \\ \frac{\partial \mbf{v}}{\partial t} &+& (\mbf{v}
\cdot \nabla)\mbf{v} +\frac{\nabla p}{\rho} + \nabla \Phi_{\rm ext} +
\nabla \Phi_{\rm sg} = 0, \label{eqn: rhov}\\
 \frac{\partial (\rho E)}{\partial t} &+& \nabla \cdot 
[(\rho E+p)\mbf{v}] = 
\rho \Gamma_{\rm UV} + \Gamma_{\star} - \rho^2 \Lambda(T_g), \label{eqn: en}\\ \nabla^2
\Phi_{\rm sg} &=& 4 \pi G \rho, \label{eqn: poi} 
\end{eqnarray}
 where, $\rho,p,\mbf{v}$ are the gas density, pressure, velocity of
the gas. The specific total energy $E \equiv |\mbf{v}|^2/2+
p/(\gamma -1)\rho$, with $\gamma= 5/3$.  

A cooling function
$\Lambda(T_g) $ $(10 < T_g < 10^8 {\rm K})$ with solar metallicity,
photo-electric heating by dust ($\Gamma_{\rm UV}$) 
and a uniform UV radiation field ($G_0 = 1.0$) are
assumed.
%$ \Gamma_{\rm UV} = 5 \times 10^{-25} \varepsilon   
%G_0 \, {\rm ergs \: s}^{-1}, $
%where $G_0$ is the incident FUV field normalized to the local
%	 interstellar value.
The hydrodynamic scheme is AUSM
 (Advection Upstream Splitting Method) \citep{liou93}
with MUSCL (Monotone  Upstream-centered Schemes for Conservation Laws).
The Poisson equation is solved to calculate self-gravity of the gas
using the Fast Fourier Transform (FFT).
%In order to calculate the isolated gravitational potential of the gas, 
%FFT is performed for a working region of $1024\times 1024 \times 128$ 
%grid points.
%We adopt implicit time integration for the cooling term.
%We set the minimum temperature, for which the Jeans instability 
%can be resolved with a grid size $\Delta$, namely
%$T_{\rm min} = 140$ K ($\rho/10 M_\odot$ pc$^{-3}$).
%The initial condition is an axisymmetric and rotationally supported
%thin disk with a uniform density.
%We run four models with different $\rho_i$ (see Table 1).
%Random density and temperature fluctuations are added to the
%initial disk. These fluctuations are less than 1 \% of the unperturbed
%values and have an approximately white noise distribution. The initial
%temperature is set to $10^4$ K over the whole region. 
%At the boundaries, all physical quantities remain at their initial
%values during the calculations. ctj

The external potential is given by 
$\Phi_{\rm ext} (R, \phi, z) \equiv \Phi_0 + \Phi_1 + \Phi_2$:
\bea
\Phi_0(R,z) &\equiv& a v_a^2 (27/4)^{1/2}(R^2+a^2+z^2)^{-1/2}, \\
\Phi_1(R,z) &\equiv& a v_b^2 (27/4)^{1/2}(R^2+b^2+z^2)^{-1/2}, \\
\Phi_2(R,\phi,z) &\equiv& \varepsilon_0 {b R^2 \Phi_1}/{(R^2 + b^2
+z^2)^{3/2} } \nonumber \\ &\times& \cos[2\phi + 2\cot i \cdot \ln (R/R_0)], 
\eea
where $i$ is the pitch angle ($i =$ 10$^\circ$) of 
the spiral potential,
and $R_0$ is an arbitrary constant, and $a=0.2$ kpc, $b=2.5$ kpc, $v_a =
v_b = 150$ km s$^{-1}$, and $\varepsilon_0 = 0.1$.
%Although the pattern speed of the spiral potential has only 
%negligible effects on the results, 
The pattern speed of the spiral potential is 
assumed $\Omega_p = 30$ km s$^{-1}$ kpc$^{-1}$. 
For comparison, we also run models with $\Omega_p = 15$ and 60 km s$^{-1}$
kpc$^{-1}$
and $\varepsilon_0 = 0.05$. 
However, we find no essential differences
on the conclusions (see also discussion in \S 4).

%Energy feedback from supernova explosions is implemented. Massive `stars'
%are formed in a grid-cell with 
%a dense ($n > 600$ cm$^{-3}$) and cold ($T_g < 100$ K).
%The stars are orbiting as massless particles in the potential, 
%and once they reach the end of their lives ($\sim 10^{6-7}$ yr), 
%they are exploded as type-II supernovae, and energy $10^{51}$ ergs
%are injected as a form of thermal energy into a grid-cell where the
% particle is located. Non-spherical propagation of blast waves
%in the multi-phase ISM is then followed in the hydrodynamic simulations
%\citep[see more details in][]{wada01}.

We take into account two feedback effects of massive stars on the gas
dynamics namely: stellar winds and supernova explosions ($\Gamma_\star$). We first
identify cells which satisfy criteria for star formation. The criteria
for each cell where star formation is allowed are: 1) The gas
density is greater than a threshold, i.e. $(\rho_g)_{i,j} > \rho_c$; 2) The temperature is less than the
critical temperature, i.e. $(T_g)_{i,j} < T_c$, and 3) The criteria 1) and
2) are satisfied for $10^5$ yr.  Here we have chosen $T_c
= 100$ K and $\rho_c = 600$ cm$^{-3}$.  Assuming the
Salpeter Initial Mass Function  with $m_u = 120 M_\odot$ and $m_l
= 0.2 M_\odot$, we create test particles representing massive stars
($>8 M_\odot$).  Typically a few massive stars are replaced by one
test particle. The initial velocity of each test particles is taken to
be the same as its parent gas.  The kinematics of the test particles
in the external potential and the self-gravitational potential of the
gas are followed by the second-order time-integration method.  The
stars (test particles) inject energy due to stellar winds to the cells
where the stars are located during their lifetime which is
approximately$\sim 10^7$ yr \citep{leith92}. 
% The heating rate used is given by
%Leitherer, Robert, \& Drissen (1992), in which solar metallicity is
%assumed. 
When the star explodes as a supernova an energy of $10^{51}$ ergs is
injected into the cell where the test particle is located. 
%The
%cooling procedure is not used for such cells, but the cells adjoining
%the supernova cell are treated normally.  
Non-spherical 3D propagation of blast waves in the multi-phase,
non-uniform ISM is then followed consistently in the hydrodynamic
simulations.

\section{RESULTS}

\begin{figure}[t]
\centering
%\caption{Same as Figure 1, but for 2D model without supernova feedback.
%The box size is 20 kpc $\times$ 20 kpc (spatial resolution is 9.8 pc).
%Pitch angle is 15$^\circ$.
%}
%\begin{center}
%	\includegraphics[scale=0.8]{er_density_two.eps}
	\includegraphics[scale=0.8]{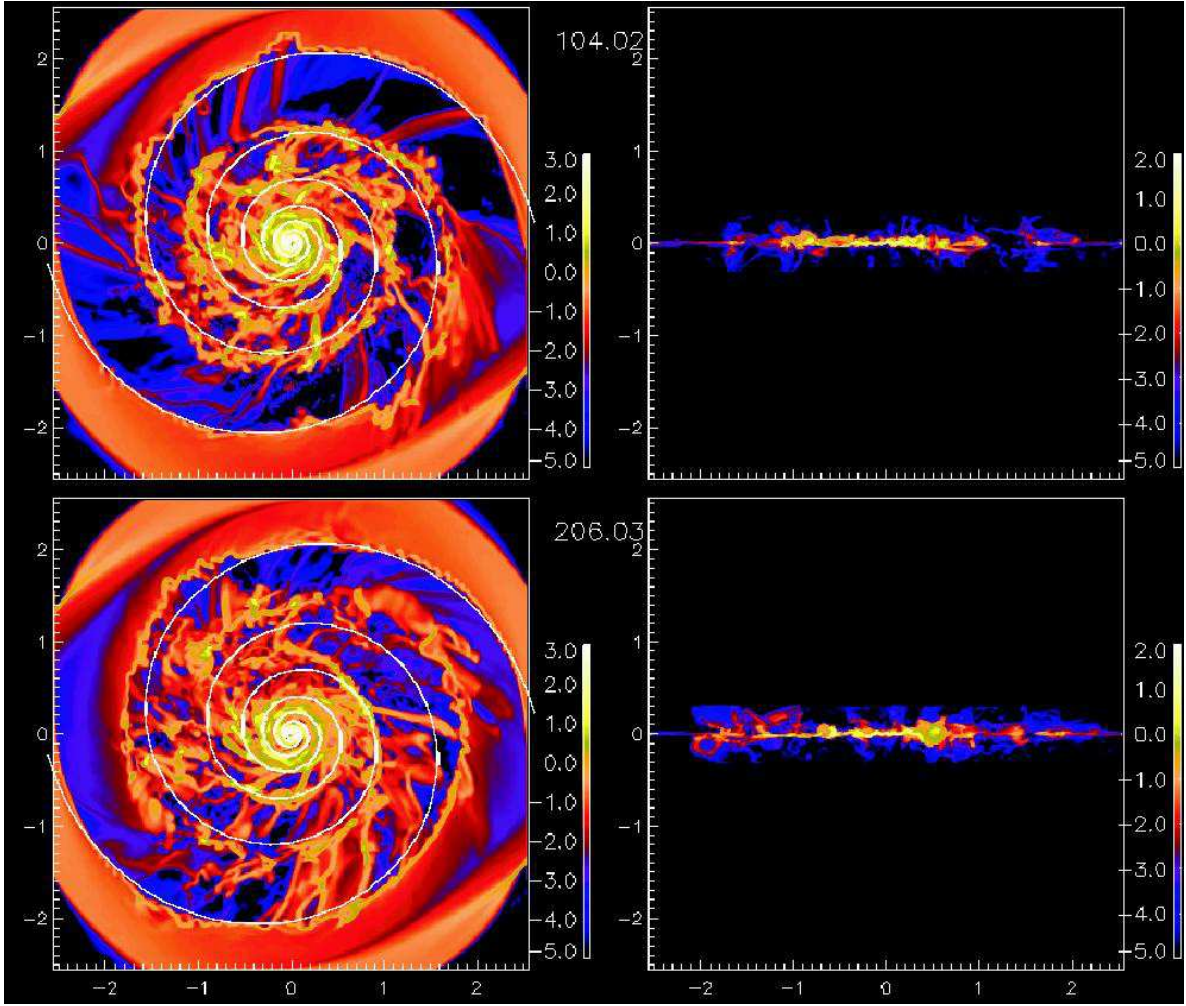}
\caption{Density distribution on a galactic plane (left panels) and x-z
 planes
(right panels) at $t=104$ and 206 Myr from 
the initial condition.
The color represents log-scaled density ($M_\odot$ pc$^{-3}$).
The unit of the scale is kpc.
(See also the Supplemental Movie for more detailed time evolution.)}
%\end{center}
\label{fig: f1}
\end{figure}

Figure \ref{fig: f1} is time evolution of 
the density field on $x-y$ and $x-z$ planes of the disk in the spiral potential.
% shows a volume-rendering representation of 
%three-dimensional structure of the density field of the gas 
%in a quasi-steady state.
High density gases form two-arm spirals, and there are also
substructures, i.e. ``spurs'', between the spirals,
as seen in real spiral galaxies \citep{elm80}.
%It has been observationally known that there are 
%radial structures, called spurs/fins/feather, 
%in a ground-design spirals.
%are oAs is observed [ref. ?, Elmegreen 1980]
Each spur is a non-steady structure, 
but spurs are always seen in our simulations (See also the {\it Supplemental Movie}).
As discussed in previous 2D simulations (Paper I)
each spur originates in a clump, whose size corresponds to observed a Giant
Molecular Cloud,  formed in the spiral arm.
A small gradient of angular momentum in the clump  is eventually
enhanced, and the clump is stretched into the inter-arm region due to galactic
rotation (see also Fig. \ref{fig: f3}). 
%In the 2D non-self-gravitating model, 
%the density fluctuation is developed in the shocked layer 
%due to the ``wiggle-instability''  \citep{wadakoda04}.
%In other words, even if self-gravity does not
%dominate in diffuse gas, the shock becomes unstable due to 
%the wiggle-instability, and spurs are formed.
In the present
model, the wiggle instability (Paper I) is
 coupled with gravitational and
thermal instabilities. 
This causes complicated distribution of the matter coupled with
the galactic rotation and tidal shear motions.
In spite of these complexities, pseudo-spiral patterns roughly associated with
the spiral potential are always present.
Energy feedback from SNe (average supernova rate is $\sim 0.2$
yr$^{-1}$) change local density and temperature of the ISM,
but it does not significantly 
affect global structures of the spirals in the galactic plane.
Note that despite the complexity, the statistical structure 
of the density field (probability distribution function of density)
is represented by a single log-normal function over a wide density range
as predicted by \citet{wada07}.
%(Fig. \ref{fig: f7}), 

\begin{figure}
\centering
	\includegraphics[scale=0.75]{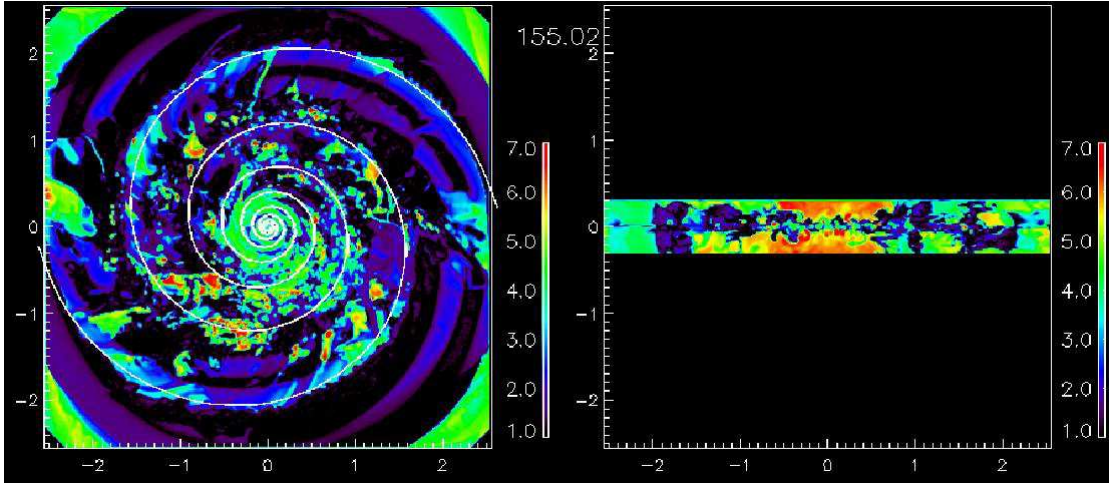}
\caption{Same as Fig. \ref{fig: f1}, but for temperature distribution at $t= 155$ Myr. The color represents
log-scaled temperature (K). The red regions are hot gases generated by
supernovae.}
\label{fig: f2}
\end{figure}

\begin{figure}
\centering
	\includegraphics[scale=0.5]{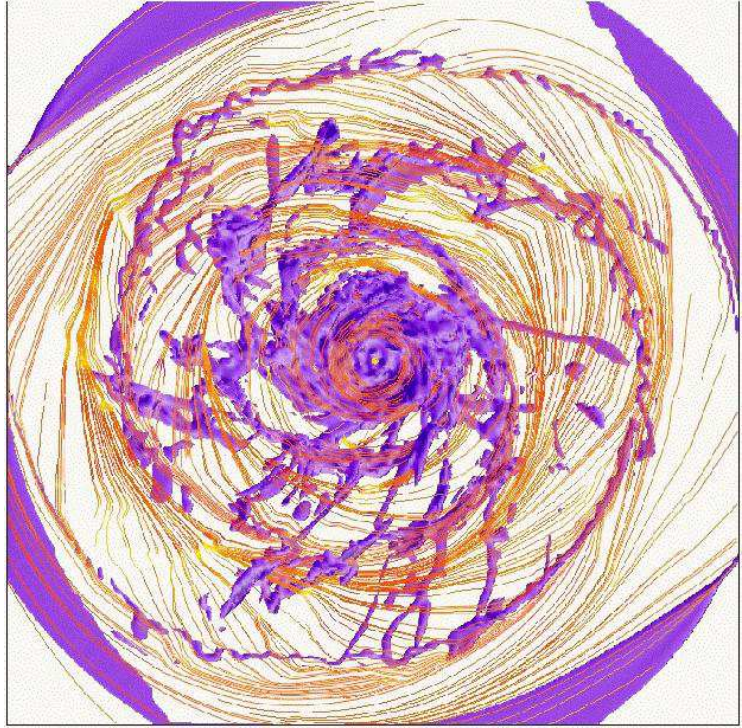}
\caption{Same as Fig. \ref{fig: f1}, but for streamlines on the galactic plane
 ($z=0$) overlaid on an iso-density surface ($\rho
>10^{-0.78} M_\odot$ pc$^{-3} \sim 10$ cm$^{-3}$) at $t= 206$ Myr. 
Note that the streamlines are 
generated by $u$ and $v$ components. The velocity field is non-steady, 
therefore the structure of the streamlines are time-dependent.}
\label{fig: f3}
\end{figure}

On the other hand, as shown in the right panels of Figs. \ref{fig: f1}
and \ref{fig: f2},
vertical distributions of density and temperature are strongly
influenced by SNe.  Cold gases ($T_g < 100$ K) as well as hot gases
($T_g \sim 10^{6-7}$ K) are blown out from the
disk plane to the halo, where warm ($T_g \simeq 10^{4-5}$ K) media
occupy most of the volume.  The cold, dense gases sometimes form
loop-like structures and fall back to the disk.  In the inner region $r
< 1$ kpc, where a supernova rate is higher than the outer region, the
halo gas is occupied mostly by the hot gases.  The
interaction between the disk and halo revealed here is a numerical
representation of a so-called ``galactic fountain'' or ``chimney''
\citep{shapiro76, normanikeuchi89}. In the original idea of the galactic fountain, hot gas is
spouted up into the halo from the disk, and it cools and subsequently
falls back to the disk.  However, our results show that not only hot
gases, but also cold, dense gases are spouted up.
The vertical structure is not steady, but in a long
time-scale, the cold dense gas is distributed within $\sim 100$ pc from
the galactic plane.  The scale height of the gas is smaller in the
central region due to the deeper gravitational potential, but
effects of a spiral potential to the vertical structure are not clearly seen.

%-- velocity structure ---

In Fig. \ref{fig: f3}, a velocity field in a disk plane ($z=0$) at $t=206$ Myr is
shown by streamlines overlaid on an iso-density
surface\footnote{The isodensity-map is a method to represent a
3D density field by a opaque surface on which the volume density is the
same. }. 
The streamlines are bent near the spiral arms, and they are spread
into the inter-arm regions.  One should note that the ``spurs'' at the
inter-arm regions are not located along the streamlines, but rather are
perpendicular to them. As already reported in 2D simulations (Paper I), the
spurs move along the galactic rotation, but they are stretched by the
spread flow. This implies that the observed ``spurs'' are not waves.
It is also apparent from Fig. \ref{fig: f3} that the density
and kinematic structures are highly irregular,
therefore it may not be represented by a local computational 
box with a periodic boundary.  Global simulations, in
which the whole galactic disk is calculated without assuming symmetry, 
are essential to understand the structure and evolution of the
ISM on a kpc scale.

%-- position of spiral shock ---

\begin{figure}[t]
\centering
	\includegraphics[scale=0.6]{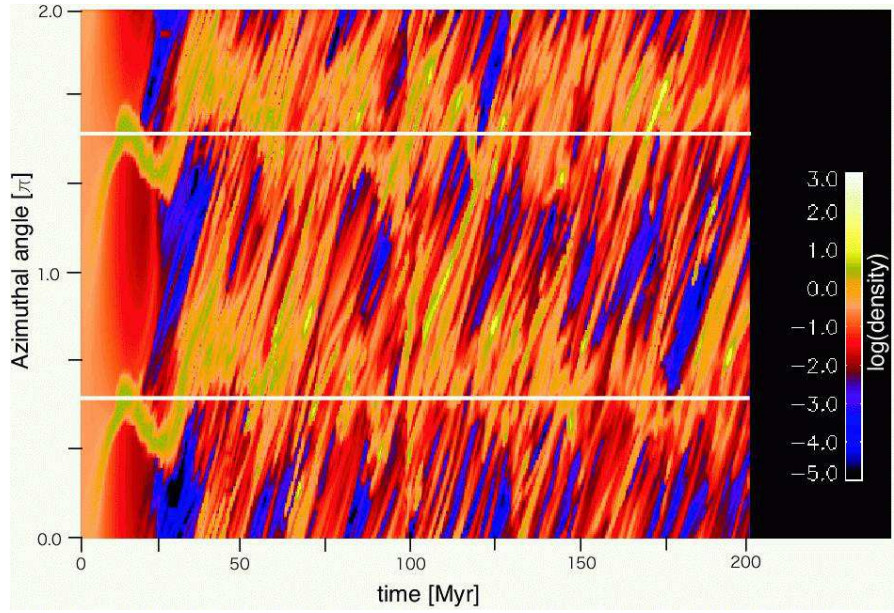}
\caption{Time evolution of azimuthal density profile at $r=1.2 $ kpc.
The color represents log-scaled density ($M_\odot$ pc$^{-3}$).
The white lines show the positions of the  minimum of the spiral potential.
The initial `shock' turns to be stochastic arms consisting of  many 
substructures after $t\simeq 60$ Myr.}
\label{fig: f4}
\end{figure}

The spiral structure seen in Figs. \ref{fig: f1} -- \ref{fig: f3} 
does not look like the typical hydrodynamic shock, in a sense that
there is no clear jump of density, and 
it is consisted of non-steady substructures.
Figure \ref{fig: f4} demonstrates long-term evolution of density at $r=1.2 $ kpc.
We see that hydrodynamic shocks are formed upstream (i.e. on the concave
side of the spiral) of the minimum of the potential trough in the first
20 Myrs or so.  This is indeed the galactic shock solution found in the
60s for a tightly wound spiral potential. However, these shocks do not
stay at the initial positions, and move back and forth between
downstream and upstream of the potential minimum until $t \simeq 60$
Myr.  After this early oscillating phase, the ``shocks'' disappear, and
a new state emerges, characterized by its stochastic nature. The
standing pattern near the potential minimum is pseudo-spiral consisting
of many clumps or parts of filaments.  Each inclined narrow
pattern seen in Fig. \ref{fig: f4} represents a dense region orbiting circularly
around the galactic center.  The trajectory can be followed between the
two spirals, i.e. for about $1/4\sim 1/2$ of the galactic rotation. 
This means that the substructures, i.e. spurs,
in the inter-arm regions are linked with the non-steady `chaotic arms'.
 Contrary to
the classic galactic shock solution, the chaotic arms found here
 are located downstream of the potential trough on average, suggesting 
that the ISM does not behave like a smooth fluid.
Typically the width of the waves is $\Delta \phi \simeq \pi/6$.
Apparently the quasi-stable structures found here are no longer `shocks'
in a hydrodynamic sense.

%%---- stars -----

Finally, positions of massive ``stars'' overlaid on a density map
at $t=155$ Myr are shown in Fig. 6. It is clear that massive stars form clusters,
most of which are distributed near the
potential trough, mostly downstream, 
and they form spiral features.  This is reasonable
since the stars originate in cold, dense clouds. In real galaxies, the
massive star complexes should form H$_{\; \rm II}$ regions around them, and
they illuminate the spiral structures. Our results show that
the H$_{\; \rm II}$ regions are offset to the downstream of the dust
spirals, which is consistent with observations \citep{delrio98,gitt04}.
 However,
as seen in Fig. 5, the spiral arms are not uniform, and positions of
density peaks in the pseudo-arms are not uniquely defined.  Therefore, it
would be difficult to derive dynamical information such as the pattern
speed of the spiral potential and the position of corotation
from observational data, for example
the offset between B-band and I-band arms or the dust-arms
and H$_\alpha$ arms \citep{delrio98,gitt04}.

%---- Discussion ---
\section{DISCUSSIONS}

\subsection{Effect of the Pattern Speed of the Potential}
%--- resonance and spiral strength ---
The density structure of the ISM in spiral potential can be affected by
two free-parameters here: the pattern speed and strength of the spiral
potential.  As mentioned in \S 1, bifurcation features of spiral arms
could be explained by Lindblad and ultraharmonic resonances
\citep{shu73, chark03}. This substructure of the spiral shocks
(i.e. ``branches'') is also seen in some models in \citet{shetty06}, but
we do not see clear evidence of the resonant structures in our results
regardless of presence of resonances.  Fig. \ref{fig: f6}c is density
distribution at $t = 61$ Myr of a model with the pattern speed $\Omega_p
= 60$ km s$^{-1}$ kpc$^{-1}$ \footnote{Since our disk is small and
compact (the core radius $R_{core}$ is 0.2 kpc and maximum circular
velocity is 150 km s$^{-1}$), $\Omega(R_{core}) - \kappa/2(R_{core})
\approx \Omega(R_{core})/2 \approx v_{max}/R_{core}$ is much larger than
that in the large disk models assumed in \citet{chark03,
shetty06}. Therefore, the pattern speed assumed here is larger than
those assumed in these previous papers (8.4 and 42 km s$^{-1}$
kpc$^{-1}$ in \citet{shetty06},
21.5 and 11.5 km s$^{-1}$ kpc$^{-1}$ in \citet{chark03}). }. As same as the fiducial model
(Fig. \ref{fig: f6}b), in which $\Omega_p$ = 30 km s$^{-1}$ kpc$^{-1}$,
the density field is characterized with inhomogeneous spiral arms and
spurs in the inter-arm regions. This is also the case in a model with
$\Omega_p = 15$ km s$^{-1}$ kpc$^{-1}$ (there are no resonances in the
disk except an ILR at $R=0.05$ kpc). In Fig. \ref{fig: f6}a, the
strength of the bar is half (i.e. $\varepsilon_0 = 0.05$) of the
fiducial model (Fig. \ref{fig: f6}b) with the same pattern speed. Same
as the fiducial models, the complicated spiral arms are formed, but it
is located more downstream than in the model with a stronger spiral
potential at this moment. 

The average location of the spiral arms is
probably affected by relative velocity of the gas to the
spiral potential and the depth of the potential.  For smaller relative
velocity (i.e. large $\Omega_p$) and/or a deeper potential, the gas
clouds tend to be trapped near the potential minimum. On the other hand,
for smaller pattern speeds and/or a shallower potential, the clouds
passes the potential minimum and decelerate.  As a result, density peaks
tend to be formed more downstream on average.
%One should
%note, however that as shown in Fig. \ref{fig: f4}, the positions of the
%spiral arm patterns are time-dependent.

%--- star/gas interactions ---
\subsection{For Further Reality}
In the present model, we assume a time-independent two-arm spiral
potential with a constant pitch angle, whereas the resultant morphology
of the gas is quite complicated and time-dependent.  These features are
qualitatively consistent with near-infrared and optical observations of
various types of spiral galaxies \citep{seigar98,grosb98}.  In spite of
the flocculent appearance of dust and young stars, the old populations
seen in the $K$-band are dominated by two-arm spirals.  The pitch angles
of $K$-band arms are distributed between 5 and 10$^\circ$ independent of
the Hubble type \citep{seigar98}.  The phase-offset between the
potential, young stars and dust-arms in the real galaxies could be used
for studying the physical origin of the spiral arms.  Unfortunately,
observational constraints on the offset of the spiral arms of different
components are ambiguous \citep{delrio98}, partly because the stellar
and gaseous arms are not smooth, and it is hard to trace their
positions, as is the case in our simulations. The distribution of the
gas and massive stars are far from regular, but more complicated than
the prediction of previous idealized models.
%\footnote{Note that
%\citet{chark03} claimed that flocculent structures of the gas can be
%reproduced due to ultraharmonic resonances with a quasi-static spiral
%potential.}

For further reality,
one should take into account the non-linear gravitational coupling
between the ISM and stellar potentials as well as the realistic treatment
of the ISM itself. This should be achieved in self-consistent, 
cosmological simulations of formation of spiral galaxies.
One should realize, however that even in recent simulations
\citep[e.g.][]{springel05, governato07}, 
the mass and spatial resolutions for gas are $10^{5-6} M_\odot$ and sub-kpc, 
and therefore  it is hard to represent the complicated multi-phase 
structures of the ISM in spiral galaxies as seen in the present simulations.
A much higher numerical resolution is also essential to
resolve the density waves in a stellar disk.
%Effects of the magnetic field on the chaotic spiral arms
%and on the vertical structure of
%the multi-phase ISM should be 
%also numerically investigated as advanced subjects. 

It is not clear how the magnetic fields
affect the stability of the multi-phase ISM in a spiral potential.
We would like to consider this problem 
in future 3D MHD simulations, in which
realistic cooling/heating processes as well as self-gravity of the gas
are at least considered. 

\section{CONCLUSION}

The high resolution hydrodynamic simulations, taking into account
self-gravity of the gas and realistic cooling and heating processes for
the ISM, reveal for the first time that the galactic spiral arms of the
ISM are neither hydrodynamic shock waves nor an assembly of long-lived,
bullet-like cloudlets.  The global spiral arms are consist of
complicated time-dependent substructures from which stars can be formed,
but over a long time (at least 5-6 rotational periods), they stably
exist under the influence of the spiral potential.  The pseudo-spiral in
the multi-phase ISM is robust for energy input from the supernovae,
which mainly cause the vertical non-uniform structure of cold and hot
gases.  The pattern speed of the spiral potential and its strength are
not key parameters to alter above features.

The ISM in galactic disks has often been represented by isothermal,
non-self-gravitating fluid or inelastic particles in many astrophysical
simulations.  Moreover, introducing a periodic boundary condition and
reducing spatial dimensions were usual. However, those kinds of
simplification do not necessarily represent the nature of spiral
galaxies.  Magnetic fields are not considered in the present
simulations, but the present results suggest that even if the magnetic
fields are weak, the spiral patterns can steadily exist on a global
scale. Effect of magnetic fields and other important physical processes,
such as UV radiation and chemistry, should be investigated 
based on 3D global models, as presented in this paper.

The present treatment of the multi-phase ISM could be used for direct
comparison with observations, coupled with radiative transfer
calculations for various observational probes \citep[e.g.][]{wada00,wada05}.
Fine structures of molecular gas associated with the spiral arms and in
the inter-arm regions in external galaxies, which could be compared with
the present numerical model, will be revealed by high resolution
observations, e.g. by the Atacama Large Millimeter/Submillimeter Array.

%%%%%%%%%%%%%%%%%%%%
\acknowledgments 

 The author thanks C. Norman and J. Koda for their
helpful comments. This work was supported by JSPS. The numerical
computations were carried out on Fujitsu VPP5000 at Center for
Computational Astrophysics in NAOJ.
%%%%%%%%%%%%%%%%%%%%%%%%%%%%%%%%%%%%%%%%%%%%%%%%%%%%%%%%%%%
%\newpage

%\newpage

%\begin{figure}
%	\begin{center}
%%	\includegraphics[scale=0.5]{3d_spiral2.eps}
%	\includegraphics[scale=0.5]{wada_fig1.eps}
%\caption{A volume-rendering representation of 
%three-dimensional structure of gas density in a time-independent spiral potential with a pitch angle of
%10$^\circ$ at $t= 155$ Myr from the initial uniform disk.
%The size of the box is $5.12\times 5.12 \times 0.64$ kpc.
%Relatively high-density regions
%are represented by an opaque color (more quantitatively see Fig. 2).
%}
%\end{center}
%\end{figure}

% Figure 1

%% Figure 2
%\begin{figure}[t]
%\centering
%	\includegraphics[scale=0.4]{pdfrho_er0400_onecomp.ps}
%\caption{Probability distribution of density at $t = 206$ Myr.
%The solid curve is a lognormal function with dispersion $\sigma = 1.6$
% and $\rho_0 = 10^{-2.2} M_\odot$ pc$^{-3}$.}
%\label{fig: f7}
%\end{figure}

% Figure 2

% Figure 3

% Figure 4

%\begin{figure}[t]
%\centering
%	\includegraphics[scale=0.6]{spiral_er_gas_mass.epsi}
%\caption{Evolution of gass mass fraction. 
%Solid line: cold ($T_g < 100$ K) gas and dashed line: Warm ($0.5 \times 10^4 <n
%T_g < 1.5 \times 10^4$ K) gas.}
%\label{fig: f4b}
%\end{figure}

% Figure 5

\begin{figure}
\centering
	\includegraphics[scale=0.4]{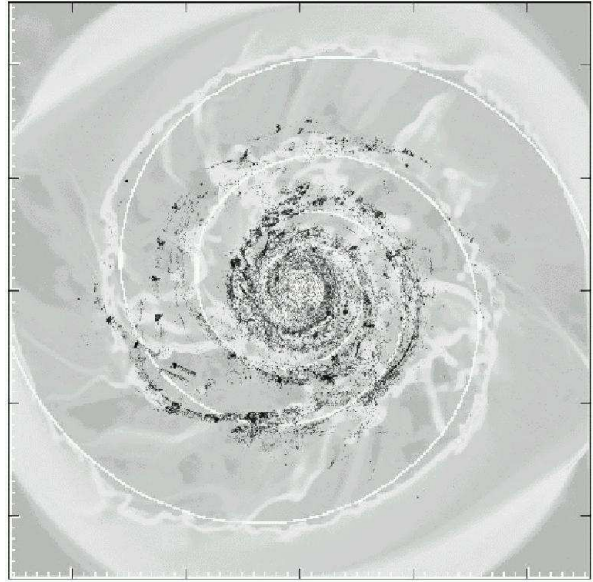}
\caption{Position of $7.9\times 10^4$ ``massive star'' particles 
overlaid on the density map at $t=155$ Myr. The size of the box is
 5.12$\times$ 5.12 kpc.}
%Density istribution of a `hot collision' model ($\gamma =
% 1.2$) at $t =13$ hours after the impact.
%}
\label{fig: f5}
\end{figure}

% Figure 6

\begin{figure}
\centering \includegraphics[scale=0.8]{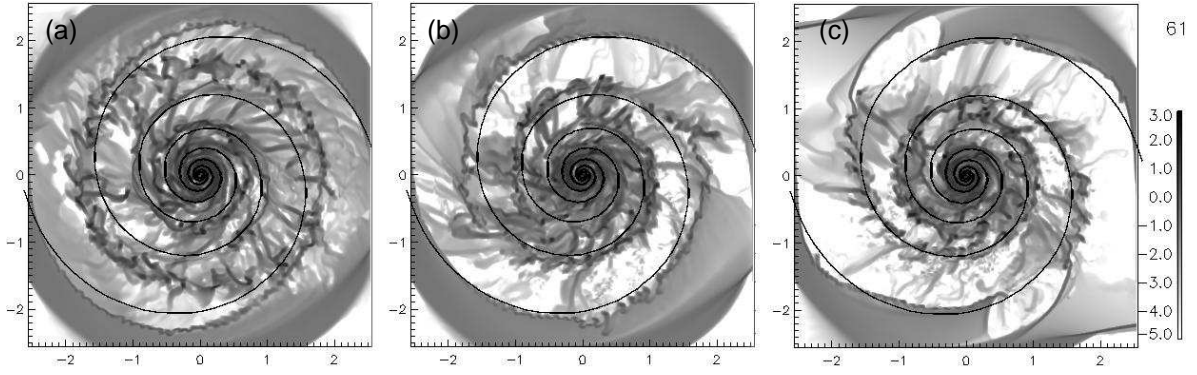} \caption{Density
distributions at $t=61$ Myr for models with (a) $\Omega_p = 30$ km
s$^{-1}$ kpc$^{-1}$ and $\epsilon = 0.05$ (b) $\Omega_p = 30$ km
s$^{-1}$ kpc$^{-1}$ and $\epsilon = 0.1$, (c) $\Omega_p = 60$ km
s$^{-1}$ kpc$^{-1}$ and $\epsilon = 0.1$.  Gray-scale represents
log-scaled density in the unit of $M_\odot$ pc$^{-3}$.  Radii of
corotation, inner Lindblad resonances, and 4:1 ultraharmonic resonance
are 2.7, 0.09, 0.74, and 1.46 kpc, respectively for (c), and 5.1, 0.05,
1.15 kpc, respectively for (a) and (b). There are no ultraharmonic
resonances in the computational region for (a) and (b). There are no
clear resonance-driven features (so-called blanches) at any resonance
radii. There is no clear sign that time-dependent inhomogeneous
substructures of spiral arms and inter-arm regions (so-called spurs) are
caused by the resononances.  }

\label{fig: f6}
\end{figure}

\end{document}